# Exact Modeling of Cardiovascular System Using Lumped Method

Omid Ghasemalizadeh[1], Mohammad Reza Mirzaee[2], Bahar Firoozabadi[2], Kamram Hassani[3]

[1] Corresponding Author: Department of Mechanical Engineering, Sharif University of Technology, Tehran, Tehran, Iran. (E-mail: alizadeh.omid@gmail.com)
[2] Department of Mechanical Engineering, Sharif University of Technology, Tehran, Tehran, Iran
[3] Chagalesh Company, Tehran, Tehran, Iran

**Abstract** - *Electrical analogy (Lumped method) is an easy way to model human cardiovascular system. In this paper Lumped method is used for simulating a complete model. It describes a 36-vessel model and cardiac system of human body with details that could show hydrodynamic parameters of cardiovascular system. Also this paper includes modeling of pulmonary, atrium, left and right ventricles with their equivalent circuits. Exact modeling of right and left ventricles pressure increases the accuracy of our simulation. In this paper we show that a calculated pressure for aorta from our complex circuit is near to measured pressure by using advanced medical instruments*
.

## 1 Introduction

Human cardiovascular diseases, which cause the majority of deaths, are the most disastrous health problems in industrialized nations. To analyze cardiovascular system and effects of diseases on it different ways are usable such as lumped model, one, two or three dimensional modeling and experimental methods. In this context different approaches were used with the goal of providing better understanding and simulation of the blood flow in the human cardiovascular system. A three-parametric model of heart muscle behaviors was introduced[1] and a modification of this model was presented later[2]. It successfully predicted force development during a time period of heart, showing deactivation of the contractile element during isotonic shortening and the apparent dependence of series stiffness in time. The study of the series elasticity of cardiac muscle was presented[3]. Afterwards, the time-varying-elastance model of the left ventricle was introduced and later the relationship between pressure-volume area and cardiac oxygen consumption was described[4]. Finite element methods are commonly used to simulate the left ventricular way of operation. The first computer models describing the arterial system were presented in the sixties. They constructed a multi-branched model of the systemic arterial tree in a form usable for digital computer, which allowed simulation of different physiological and pathological conditions. This model was extended in detail later[5]. Also, blood flow through sites of particular interest of the arterial tree such as anastomoses, stenoses and bifurcations can be described with finite element method simulation. The mathematical analysis of the whole human cardiovascular system remains as a complicated task and for that reason models are simplified with respect to particular parts of interest. A pulsatile-flow model of the left heart and two-segment aorta were constructed and the changes in flow work investigated. In addition, Time-tension index and stroke volume, which is produced by an intra aortic balloon pump, were inspected. Later an open-loop lumped parameter model was developed and the sensitivity of an intra aortic balloon pump to timing, rate of inflation, balloon placement, and stroke volume was studied[6]. The model later resulted in good outcomes for a short-term interaction between the assist device and the circulation[7]. The model of an assisted failing canine circulation was expressed from which cardiac oxygen supply and consumption could be calculated[8]. The impact of drugs on the circulation was studied later in a 19-compartment model. Also, the use of computer-aided design of control systems was improved to produce a simple electrical analog of the human cardiovascular system. It has been used to simulate steady-state circulatory conditions with transients introduced by varying the peripheral resistance[9]. A non-linear computer model for pressure and flow propagation in the human arterial system was drived[10]. The model with 55 arterial segments is based on one-dimensional flow equation and simulates different homodynamic effects on blood flow. The simulation of steady state and transient phenomena using the electronic circuit has also been handled but the arteries have not been included in their model widely[11]. An electrical model which focused on the vessel beds was presented by Young[12]. One-dimensional axisymmetric Navier-Stokes equations for time dependent blood flow in a rigid vessel had been used to derive lumped models relating flow and pressure[13].

This paper describes simulation of the cardiovascular system using a complex electronic circuit. In this study we have taken a slightly different approach to the modeling of the system and tried to advance existing electrical models by increasing the number of influential segments and parameters. The model consists of more than 60 segments representing the arterial system. Compared to previous studies, the arterial system of this model is more detailed. Therefore, normal and abnormal performance of arterial system can be investigated.

Here is a 36-vessel body tree which is utilized in this research.

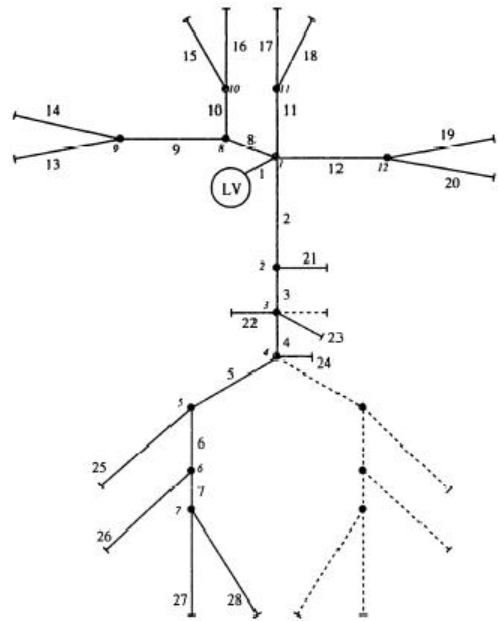

**Fig1 – 36 Vessels Body Tree**

## 2 Modeling Principals

In our model, every blood vessel, atrium, ventricle and set of all capillaries and arterioles have been presented by a block consisting of a resistor, an inducer and a capacitor.

Voltage, current, charge, resistance and capacitance in the electronic circuit are respectively equivalent to blood pressure, blood flow, volume, resistance and compliance in the cardiovascular system. Ground potential (reference for voltage measurements) is assumed to be zero as usual. The correlation between electrical characteristics of the system and their mechanical counterparts are as follow:

0.01ml/Pa = 1 µF (compliance capacitance)
1 $Pa.s^2$/ml = 1 µH (inertia     inductor)
1 Pa.s/ml = 1 kΩ (resistance - resistance)
1mmHg = 1 volt (pressure     voltage)
133416 ml = 1A (volume     charge)

Following formulas is taken for simulation[2].

Blood vessel resistance (R), depending on blood viscosity and vessel diameter, is simulated by resistors:

$$R = \frac{8l\pi\mu}{A^2} \quad (1)$$

Where μ is blood viscosity, l and A are in respect length and cross section area of each artery segment.

This simulation has considered because blood viscosity will cause resistance against Blood flow crossing.

The blood inertia (L) is simulated by inductors:

$$L = \frac{9l\rho}{4A} \quad (2)$$

Where ρ is blood density.

Reason of this consideration is variability of flow acceleration in pulsatile blood flow, so an inductor can model inertia of blood flow very clearly.

The vessel compliance (C) is considered using capacitors:

$$C = \frac{3l\pi r^3}{2Eh} \quad (3)$$

Where r, E, h are in respect artery radius, Elasticity module and thickness of arteries.

For the reason of this simulation, it should be noted that by passing blood thorough vessels, the vessels would be expanded or contracted, so they can keep blood or release it and this is exactly like what a capacitor does. By these statements each vessel is modeled by some compartments, which includes one resistance, one capacitor, and one inductor. Quantities of Compartment's elements are easily achievable by using equations 1, 2, and 3. Using more compartments to model arteries has this advantage to reach more accuracy.

Atriums are simulated as part of the venous system without any contraction. Atriums are modeled as a resistor-capacitor segment. Ventricles are simulated as a section of blood vessels in which its resting capacitance (diastole) can be decreased (systole) and then returned to its first condition.

Essentially, energy of systolic contraction of left and right ventricles is modeled quite exactly from biological graph sources.

Different heart shutters are modeled by using appropriate diodes, because shutters like diodes cross the flow in one direction.

Also, bifurcations are important cases to have accurate modeling for simulating these parts of cardiovascular system a special method has been used[2].

## 2.1 Circuit Description:

The equivalent circuit of the cardiovascular system is shown in Appendix. In this simulation, cardiac output and aortic pressure should be 100 (ml/s) and 120-68 mmHg[14]. This was exactly calculated in our circuit.

The equivalent circuit can be subdivided into three subdivided parts:

- heart (ventricles, atriums, pulmonaries and shutters)

The left atrium and ventricle are represented by two capacitors 101 μF and 25 μF. Also right atrium and ventricle are modeled by two capacitors 216.45 μF and 150 μF [2]. Using of suitable diodes was very important matter in this circuit. The aortic, mitral, tricuspid and pulmonary valve are simulated with ideal diodes: SD41, SD41, 120NQ045 and QSCH5545/- 55C. Using these types of diodes has helped the circuit to show the operation of the cardiovascular system properly. Also an exact model of ventricles pressure has been used as the supply of power. The simulated and exact pressure graphs have been compared in fig2 and fig3 for right ventricle and fig4 and fig 5 for left ventricle.

It is obvious that left and right ventricles pressure will change, by turn, between 120-11 volt (mmHg) and 29-7 volt (mmHg).

Also arterioles have been simulated as one separated compartment with resistance of 72kΩ and capacitance of 1.4μF. The capillaries and veins are also modeled by one and two segments and their quantity is achievable by referring to the Appendix.

- upper part of body (hands and carotids)
- downer part of body (thoracic aorta and feet)

In these to cases elements could be determined by using equation 1, 2 and 3.

Calculated quantities for vessels elements are shown in table 1.

The most important point in making these two sets of elements is that the vessels which are end to capillaries should be connected to the arterioles before heart circuit.

The generated current of suppliers distributes to the left ventricle, aorta and upper body arteries. It continues its path toward the body arteries such as thoracic and femoral. Then the current passes the arterioles, capillaries and veins and enters to the right atrium.

Another amplifier (right ventricle), Generates the required current for circulation in the pulmonary arteries and veins and the whole current enters the left atrium finally. Model is capable of showing pressure signals for different arteries throughout ascending aorta to femoral triggered by initial systolic contraction. The pressure (voltage) and current (Flow rate) graphs can be obtained from the different points of the circuit easily.

It shall be noted that there is no leakage of current in the system and the output voltage is proportional to the input voltage. Total capacitance of the system is one of the important determinants of its functions. The ratios of capacitances determine the distribution of total charge in the circuit. The ratio between total systemic arterial and systemic venous capacitance is about 0.01.

The ratio between total systemic and pulmonary is 0.15.

These ratios are similar to those reported for the human cardiovascular system.

The computed values of Elements are shown in table 1. Where $\rho$ is 1050 kg/m$^3$ and $\mu$ is 0.0035 kg/m.s.

Also, thickness of artery wall (h) is obtained from physiological text[14].

## 3 Results

The results of simulation performed for a heart frequency of 1.219Hz, are given as follow.

Fig2 shows the calculated pressure graph of right ventricle varying between 29–7 mmHg (volt). The actual graph of right ventricle confirms this simulation result in accordance with physiological data of reference[14] (Fig3).

The calculated pressure-time graph of left ventricle is shown in Fig4, where the waveform varies between 120–11 mmHg (volt). The results are in complete agreement with experimental observation of physiological Text[14] (Fig5). The waveform starts from 11 mmHg and the peak is in 120 mmHg.

The calculated pressure changes of ascending aorta artery are shown in Fig6. This graph shows that aorta pressure varies between 120–68 mmHg (volt) (systole-diastole) and the results are in exact agreement with physiological article[14] (Fig7). Even two peaks in aorta pressure are earned as the same of real measurement.

In the equivalent circuit, cardiac output can be strongly increased by increasing the value of capacitor simulating left ventricle.

At last it should be noted that it seems that use of this equivalent electronic circuit of cardiovascular system is useful for studying of whole cardiovascular system and is a useful tool in teaching physiology and pathology for students. Also different cardiovascular pathologies such as arterial or cardiac abnormities can be studied with changing model's parameters.

## 4 Conclusion

At last it should be noted that use of this equivalent electronic circuit of cardiovascular system with so many details which is simulated is so useful for studying of whole cardiovascular system and its abnormalities such as obstructions. Also, different cardiovascular pathologies such as arterial or cardiac abnormities can be studied with changing model's parameters.

At the end it should be said that we can increase the accuracy of results by adding more and more compartments to our circuit.

**Table 1 – Calculated Values for Elements of Circuit from artery parameters**

| Vessel Number | Vessel Name | A (cm^2) | l (cm) | h (cm) | r (cm) | E (Mpa) | R (KΩ) | C (µF) | L (µH) |
|---|---|---|---|---|---|---|---|---|---|
| 1 | Ascending Aorta | 6.60507 | 5.5 | 0.163 | 1.449987 | 0.4 | 0.01109 | 1.211848 | 0.196724 |
| 2 | Thoracic Aorta | 3.59701 | 18.5 | 0.124 | 1.07003 | 0.4 | 0.125776 | 2.15337 | 1.215072 |
| 3 | Abdominal Aorta | 2.378 | 4.3 | 0.11 | 0.870024 | 0.4 | 0.066889 | 0.303284 | 0.427197 |
| 4 | Abdominal Aorta | 1.021 | 9.6 | 0.08 | 0.570083 | 0.4 | 0.810079 | 0.261925 | 2.221352 |
| 5 | Common iliac | 0.849 | 19.2 | 0.076 | 0.519851 | 0.4 | 2.343114 | 0.418125 | 5.342756 |
| 6 | Femoral Artery | 0.181 | 43.2 | 0.048 | 0.240029 | 0.8 | 115.9936 | 0.073314 | 56.38674 |
| 7 | Anterior Tibial Artery | 0.053 | 1.5 | 0.035 | 0.129886 | 1.6 | 46.97287 | 0.000277 | 6.686321 |
| 8 | Brachiocephalic | 1.20798 | 2.4 | 0.085 | 0.62009 | 0.4 | 0.144677 | 0.079312 | 0.469379 |
| 9 | R Brachial | 0.50298 | 41 | 0.064 | 0.40013 | 0.4 | 14.25568 | 0.483491 | 19.25768 |
| 10 | R Common Carotid | 0.50298 | 16.8 | 0.064 | 0.40013 | 0.4 | 5.841352 | 0.198113 | 7.890951 |
| 11 | L Common Carotid | 0.50298 | 11 | 0.064 | 0.40013 | 0.4 | 3.824695 | 0.129717 | 5.166694 |
| 12 | L Brachial | 0.55399 | 44.4 | 0.066 | 0.41993 | 0.4 | 12.72576 | 0.586882 | 18.9344 |
| 13 | R Radial | 0.08 | 23.2 | 0.043 | 0.15958 | 0.8 | 318.8472 | 0.012915 | 68.50991 |
| 14 | R Ulnar | 0.13901 | 22.9 | 0.047 | 0.21035 | 0.8 | 104.2495 | 0.026713 | 38.91999 |
| 15 | R External Carotid | 0.196 | 11.3 | 0.049 | 0.249777 | 0.8 | 25.87461 | 0.021169 | 13.62054 |
| 16 | R Internal Carotid | 0.283 | 17.2 | 0.054 | 0.300136 | 0.8 | 18.89136 | 0.050727 | 14.35866 |
| 17 | L Internal Carotid | 0.283 | 17.2 | 0.054 | 0.300136 | 0.8 | 18.89136 | 0.050727 | 14.35866 |
| 18 | L External Carotid | 0.196 | 11.3 | 0.049 | 0.249777 | 0.8 | 25.87461 | 0.021169 | 13.62054 |
| 19 | L Radial | 0.13901 | 23.2 | 0.047 | 0.21035 | 0.8 | 105.6152 | 0.027063 | 39.42986 |
| 20 | L Ulnar | 0.08 | 22.9 | 0.043 | 0.15958 | 0.8 | 314.7242 | 0.012748 | 67.624 |
| 21 | Coeliac | 0.478 | 1 | 0.064 | 0.390067 | 0.4 | 0.384992 | 0.010925 | 0.494247 |
| 22 | Renal | 0.212 | 2.7 | 0.049 | 0.259773 | 0.4 | 5.284447 | 0.01138 | 3.008844 |
| 23 | Sup Mesenteric | 0.581 | 5.4 | 0.066 | 0.430044 | 0.4 | 1.407178 | 0.07666 | 2.195783 |
| 24 | Inf Mesenteric | 0.08 | 4.5 | 0.043 | 0.159577 | 0.4 | 61.85005 | 0.00501 | 13.28906 |
| 25 | Profundis | 0.166 | 12.1 | 0.047 | 0.229868 | 1.6 | 38.62573 | 0.00921 | 17.22063 |
| 26 | Post Tibial | 0.102 | 30.6 | 0.043 | 0.180188 | 1.6 | 258.7192 | 0.012262 | 70.875 |
| 27 | Ant Tibial | 0.031 | 29.5 | 0.03 | 0.099336 | 1.6 | 2700.264 | 0.002839 | 224.8185 |
| 28 | Proneal | 0.053 | 31.3 | 0.035 | 0.129886 | 1.6 | 980.1671 | 0.005771 | 139.5212 |

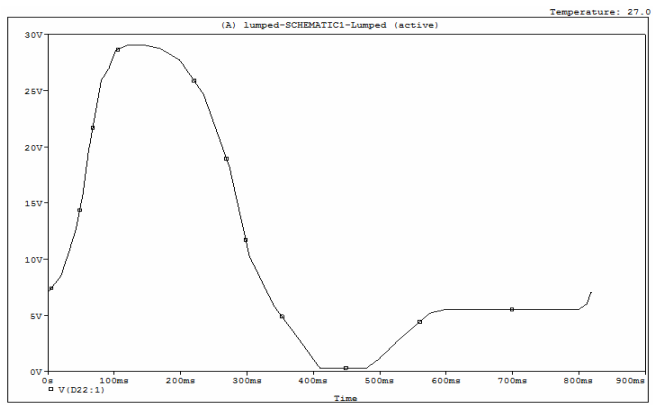

**Fig2 – Calculated Pressure Graph for Right Ventricle**

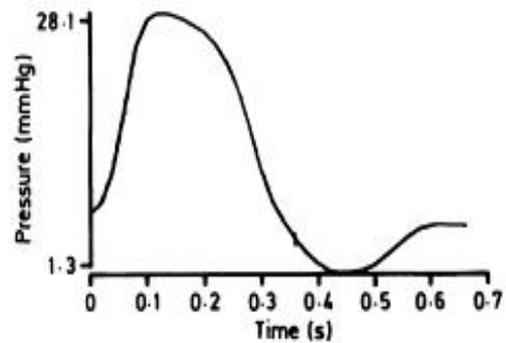

**Fig3 – Real Pressure Graph for Right Ventricle**

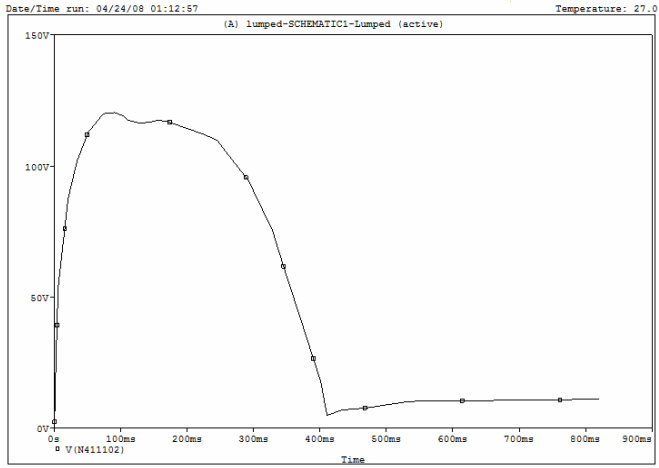 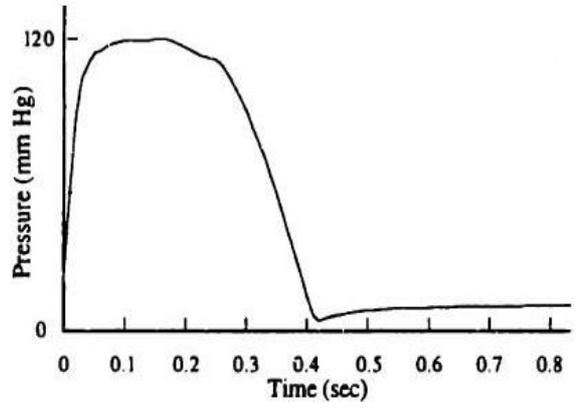

**Fig4 –Calculated Pressure Graph for Left Ventricle**    **Fig5 – Real Pressure Graph for Left Ventricle**

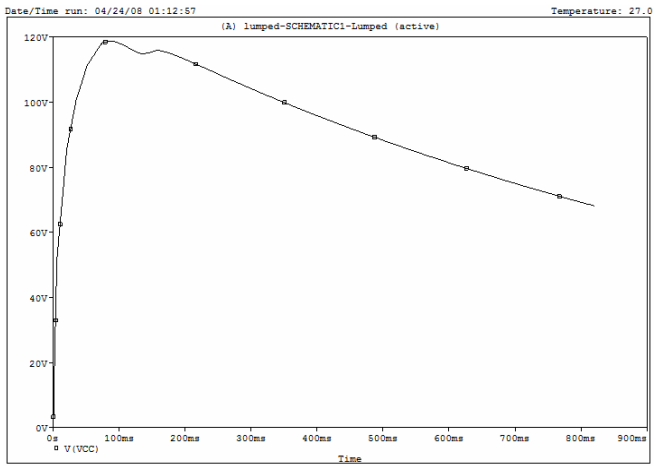 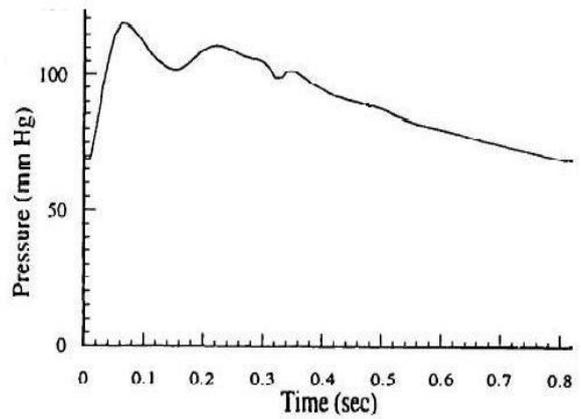

**Fig6 – Calculated Pressure Graph for Ascending Aorta**    **Fig7 – Real Pressure Graph for Ascending Aorta**

# APPENDIX

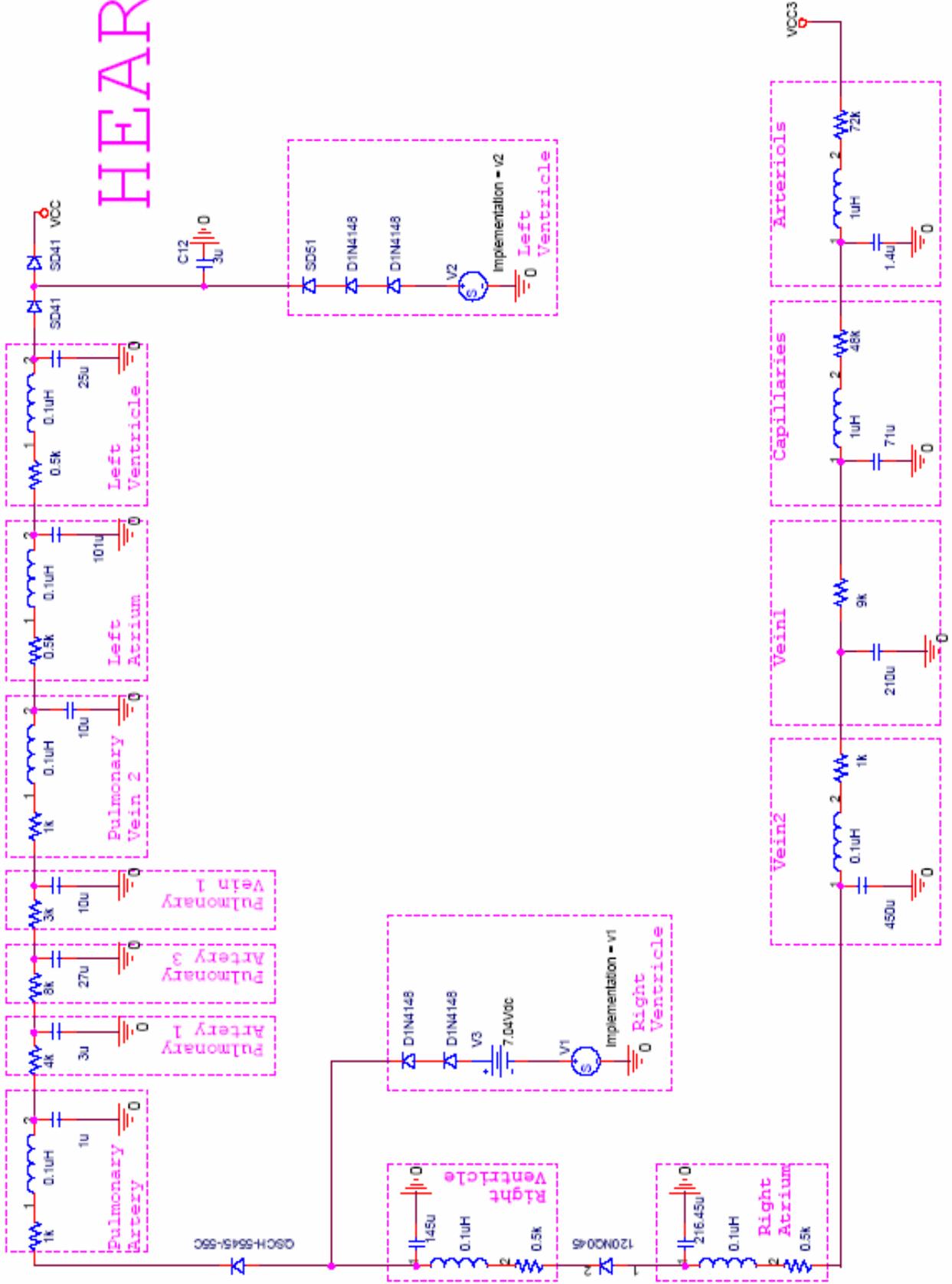

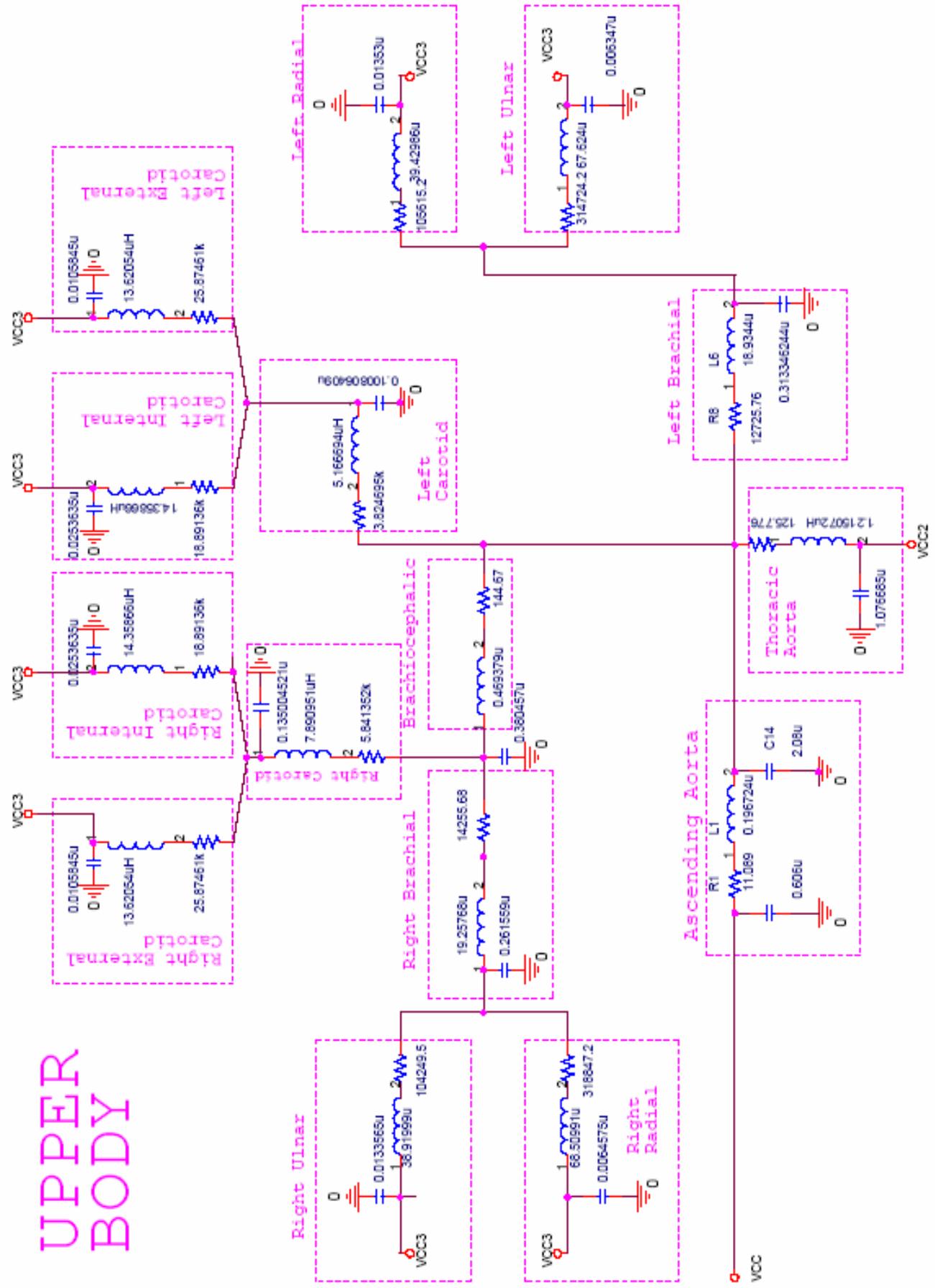

# DOWNER BODY